\documentclass[aps,pra,reprint,superscriptaddress,longbibliography]{revtex4-2}
\usepackage{multirow}
\usepackage{graphicx}
\usepackage{enumerate}
\usepackage{enumitem}
\usepackage{amsmath}
\usepackage{amssymb}
\usepackage{stix}
\usepackage{mathtools}
\usepackage{wasysym}
\usepackage{fontawesome}

\newtheorem{theorem}{Theorem}[section]

\usepackage{float}

\begin{document}

\def\crta{\vrule height1.41ex depth-1.27ex width0.34em}
\def\dj{d\kern-0.36em\crta}
\def\Crta{\vrule height1ex depth-0.86ex width0.4em}
\def\Dj{D\kern-0.73em\Crta\kern0.33em}
\dimen0=\hsize \dimen1=\hsize \advance\dimen1 by 40pt

\title{Comment on ``Optimal conversion of
  {K}ochen-{S}pecker sets into bipartite
  perfect quantum strategies''}

\author{Mladen Pavi\v ci\'c}
\email{mpavicic@irb.hr}

\affiliation{Center of Excellence for Advanced Materials 
and Sensing Devices (CEMS), Photonics and Quantum Optics Unit,\\ 
Ru{\dj}er Bo\v skovi\'c Institute, Zagreb, Croatia.}

\begin{abstract}
  A recent paper of Trandafir and Cabello
  [Phys.~Rev.~A, {\bf 111}, 022408 (2025)] contains a number of
  errors, inconsistencies, and inefficiencies. They are too
  numerous to be listed here, so we identify and discuss them
  in the main body of the comment. 
  \end{abstract}

%\pacs{03.67.Dd, 03.67.Ac, 42.50.Ex}
 \maketitle

\section{\label{esc:intro}Introduction }

Trandafir and Cabello \cite{cabello-25} introduced an algorithm
that identifies the bipartite perfect quantum strategies for a
given Kochen-Specker (KS) set in a nonlocal game. They apply it
to a number of historical KS sets in dimensions 3-8. In doing
so they were inefficient, made a number of errors and created
some inconsistencies. As a service to the reader we pinpoint them
below. We also draw reader's attention to the fact that there
exist programs for automated generation of any KS or non-KS set
in a language which enables user-friendly processing of obtained
sets and their coordinatizations.

Trandafir and Cabello \cite[Sec.~I.D.]{cabello-25} start with a
definition of a KS set via the KS theorem by means of projectors
but later on keep to vectors. So, we start with definitions of
KS and non-KS sets with the help of vectors. Notice that they
call our KS sets---``extended'' KS sets and our non-KS sets just
KS sets. 

 \smallskip
\begin{theorem}\label{theorem:KS}
  {\bf KS (non-KS)}
  {\rm \cite{zimba-penrose,renner-wolf-04,pmmm05a,pavicic-quantum-23,pavicic-entropy-25}.}
  In ${\cal H}^n$, $n\ge 3$, there are sets, called KS
  (non-KS) sets, of $n$-tuples ($m$-tuples, $2\le m\le n$)
  of mutually orthogonal vectors to which it is impossible
  to assign $1$s and $0$s so that

{\parindent=0pt
(i)  No two orthogonal vectors are both assigned the value $1$;

(ii) In no $n$-tuple ($m$-tuple) of mutually orthogonal
  vectors, all of the vectors are assigned the value $0$.}
\end{theorem}

Both KS and non-KS sets are contextual. 

\smallskip
We shall make use of the hypergraph language
\cite{pmmm05a,pavicic-quantum-23,pavicic-entropy-25,budroni-cabello-rmp-22}
since it is more efficient to present sets and follow
presentation.

We encode vectors with the help of ASCII characters
by one of the following 90 characters:
{{\tt 1 2 \dots\ 9 A B \dots\ Z a b
 \dots\ z !\ " \#} {\$} \% \& ' ( ) * - / : ; \textless\ =
\textgreater\ ? @ [ {$\backslash$} ] \^{} \_ {`} {\{}
{\textbar} \} $\sim$}\ \ \cite{pmmm05a}. A 91st character
`+', is used for the following purpose: when all aforementioned 
characters are exhausted, we reuse them prefixed by `+', then
again by `++', and so on. Vector 10 in \cite{cabello-25}
corresponds to {\tt A}, vector 11 to {\tt B}, and so on.  

We encode $m$-tuples ($2\le m\le n$) of orthogonal $n$-dim
vectors by strings of their codes without spaces. For instance,
3Ag+\&+++:x\_9 is an 8-tuple in an 8-dim space.

A KS (non-KS) set is organized as a string containing
$m$-tuple-substrings separated by commas (`,') and ending with
a period (`.'). There is no limit on the size of a set.

A KS (non-KS) set with $k$ vectors and $l$ $\,$$m$-tuples is
called a $k$-$l$ KS (non-KS) set. 

A KS (non-KS) set is {\em critical\/} if removing any of its
$n$($m$)-tuples turns it into a noncontextual set for which both
{\em(i)} and {\em(ii)} from Theorem \ref{theorem:KS} are satisfied.

Graphically, vectors in a set can be represented by means of
dots and $m$-tuples of mutually orthogonal vectors as lines
or curves passing through the dots.

\section{\label{sec:errors} Efficiency, consistency,
  and errors of \cite{cabello-25}}

In the following sections $\boxed{\ne}$ means ``wrong,''
 \faUnlink\  means inconsistent, and
\faHourglassHalf\ means inefficient. We first
cite some sentences from \cite{cabello-25} and then explain
why the results they refer to are wrong, or inconsistent,
or inefficient.

Notice that statements by Trandafir and Cabello I found to be wrong
and denoted by $\boxed{\ne}$ refer to sets that have only full triples
(each triple has three vectors) and that none of the conditions of the
KS theorem \ref{theorem:KS} are violated. Hence, the considered sets
are neither KS nor non-KS nor “extended” KS sets. They all allow 0-1
assignments and are therefore noncontextual.

\subsection{\label{subsec:4dim01} IV A,B,C and TABLE III of
  \cite{cabello-25} (\faHourglassHalf\ )}

\faHourglassHalf\ ``TABLE III. P-24, K-20, CEG-18  and their
optimal\break B-KSs..'' \cite[p.~7]{cabello-25}

The authors refer to three historical papers and adopt a notation
that creates difficulties while user evaluation. Instead, we suggest
a language whose automated outputs enable subsequent feeding of
programs for finding optimal B-KSs.

For example, Peres' P-24 follows from vector components
$\{0,\pm1\}$ as follows:

COMMAND-M: vecfind -4d -nommp -master -vgen=0,1,-1 | mmpstrip -U |
states01 -1 |  grep fails

OUTPUT: {\bf 24-24} fails (admits no {0,1} state): {\tt 125E,12NW,\break 15HK,18BE,25FG,267E,345E,34NW,3Nbd,3SUW,4Nae,\break 4RVW,67HK,6HVd,6KSa,7HUe,7KRb,8BFG,8FVb,8GUa,\break BFSe,BGRd,RVbd,SUae.}\hfil\{{\tt 1}=(0,0,0,1), {\tt 2}=(0,0,1,0), {\tt 3}=(0,0,1,1), {\tt 4}=(0,0,1,-1), {\tt 5}=(0,1,0,0), {\tt 6}=(0,1,0,1), {\tt 7}=(0,1,0,-1), {\tt 8}=(0,1,1,0), {\tt 9}=(0,1,1,1), {\tt A}=(0,1,1,-1), {\tt B}=(0,1,-1,0), {\tt C}=(0,1,-1,1), {\tt D}=(0,-1,1,1), {\tt E}=(1,0,0,0), {\tt F}=(1,0,0,1), {\tt G}=(1,0,0,-1), {\tt H}=(1,0,1,0), {\tt I}=(1,0,1,1), {\tt J}=(1,0,1,-1), {\tt K}=(1,0,-1,0), {\tt L}=(1,0,-1,1), {\tt M}=(-1,0,1,1), {\tt N}=(1,1,0,0), {\tt O}=(1,1,0,1), {\tt P}=(1,1,0,-1), {\tt Q}=(1,1,1,0), {\tt R}=(1,1,1,1), {\tt S}=(1,1,1,-1), {\tt T}=(1,1,-1,0), {\tt U}=(1,1,-1,1), {\tt V}=(1,1,-1,-1), {\tt W}=(1,-1,0,0), {\tt X}=(1,-1,0,1), {\tt Y}=(-1,1,0,1), {\tt Z}=(1,-1,1,0), {\tt a}=(1,-1,1,1), {\tt b}=(1,-1,1,-1), {\tt c}=(-1,1,1,0), {\tt d}=(1,-1,-1,1), {\tt e}=(-1,1,1,1)\}

``Six critical KS sets contained in 24-24'' \cite[p.~6]{cabello-25}
one can better obtain from \{-1,0,1\} with the following
command (after 19 sec on a PC):   

COMMAND-V: vecfind -4d -nommp -master -vgen=0,1,-1 | mmpstrip -U | states01 -1 | grep fails | sed 's/\^{}.*:: //' | sed 's/\..*\$/\./g' | states01 -1 -r6666 | sed 's/\^{}.*:: //' | shortd -G | mmpshuffle -na | vecfind -4d -nk -vgen=0,1,-1 | states01 -1 -c 

OUTPUT: \#1 {\bf 18-9} parity-passes (is critical): 1234,5678,\break 9ABC,DEC8,FGB7,HGA4,IE93,IF62,HD51. \{{\tt 1}=(0,0,0,1), {\tt 2}=(0,1,1,0),\dots---full output is given in \cite{pmmm05a}.

\#2 {\bf 20-11(a)} parity-passes (is critical): 1234,5678,9A84,\break
BC73,\dots---full output is given in \cite{pmmm05a}.

\#3 {\bf 20-11(b)} parity-passes (is critical): 1234,5674,8973,\break ABC9,DEC8,FGE6,HGB5,IJE2,KJB1,IFDE,KHAB. ({\tt 1}=(1,-1,0,0), {\tt 2}=(0,0,1,-1), {\tt 3}=(1,1,1,1), {\tt 4}=(1,1,-1,-1),\break {\tt 5}=(0,1,1,0), {\tt 6}=(1,0,0,1), {\tt 7}=(1,-1,1,-1), {\tt 8}=(0,1,0,-1), {\tt 9}=(1,0,-1,0), {\tt A}=(0,1,0,0), {\tt B}=(0,0,0,1), {\tt C}=(1,0,1,0), {\tt D}=(1,1,-1,1), {\tt E}=(-1,1,1,1), {\tt F}=(1,1,1,-1), {\tt G}=(0,1,-1,0), {\tt H}=(1,0,0,0), {\tt I}=(1,-1,1,1), {\tt J}=(1,1,0,0), {\tt K}=(0,0,1,0)\}

\#4 {\bf 22-13(a)}\dots \#5 {\bf 22-13(b)}\dots ---full
outputs are given in \cite{pmmm05a}.

\#6 {\bf 24-15}\dots ---full output is given in
\cite{pmm-2-10}.

\smallskip
It is interesting to learn that the authors of the papers arriving at
P-24, K20, and K18 (24-24, 20-11(b), and 18-9) spent years to obtain
them in the previous century, but that belongs to the history of
science. Instead, the reader of the {\em Physical Review\/} should
be informed of the most efficient way of extracting such data
especially in the form that enables her/him to further process them
further in an automated way. 

\subsection{\label{subsec:4dim02} IV D,E and TABLES IV,V
  of \cite{cabello-25} ($\boxed{\ne}$,
  \faUnlink\ , \faHourglassHalf\ )}

\faUnlink\ , \faHourglassHalf\ ``Penrose’s 40-vector
KS \dots Pen-40, is a complex KS set.'' \cite[p.~8]{cabello-25}

\smallskip 
While the authors gave the coordinatization for all the other of
their examples they did not give it for Penrose's 40-40. 
The references the authors provide do not offer an explicit
coordinatization, either. Details are given below. 

\smallskip 
$\boxed{\ne}$ ``Zimba-Penrose’s 28-vector KS set \dots is a
critical KS set formed by removing 12 implicit vectors of Pen-40.''

\smallskip 
This is wrong. The 28-14 set in question is a noncontextual set,
i.e., it is neither a KS nor a non-KS set. Also, the
coordinatization is not provided. Details are given below. 

\smallskip

When we make use of vector components $\{0,\pm1,\omega,\omega^2\}$,
where $\omega=e^{2\pi i/3}=(-1+i\sqrt{3})/2$, instead of
$\{0,\pm1\}$ in the 1st command in Sec.~\ref{subsec:4dim01},
we obtain a 364-796 master set. From it, via parallelizing the
2nd command in Sec.~\ref{subsec:4dim01} on a supercomputer we obtain 
almost all KS set the master set contains including the 40-40.

{\bf 40-40} {\tt 248N,359O,6ADc,67Ed,78Fe,89Ba,BDHW,BEKZ,\break 1BLV,9ACb,4EOY,3GKS,4GHT,7HRb,5HIU,8ISc,CEIX,\break 1IJQ,2JKR,9JTd,DFJY,256L,LSYb,LTXe,137M,2CMW,\break MTZc,MUYa,3DNX,14AP,5FPZ,PRXa,PSWd,6GQa,NQZb,\break OQWe,AKUe,CFGV,NUVd,ORVc.} \{{\tt 1}=(0,0,0,1), {\tt 2}=(0,1,1,1), {\tt 3}=(0,0,1,0), {\tt 4}=(1,-1,1,0), {\tt 5}=($\omega$,1,0,-1), {\tt 6}=($\omega$,0,-1,1), {\tt 7}=(0,1,0,0), {\tt 8}=(1,0,-1,1), {\tt 9}=(1,$\omega$,0,-1), {\tt A}=($\omega^2$,-1,$\omega$,0),\break {\tt B}=(-1,$\omega$,-1,0), {\tt C}=(0,1,$\omega$,$\omega^2$), {\tt D}=($\omega$,$\omega^2$,0,-1), {\tt E}=(1,0,-1,$\omega$),\break {\tt F}=($\omega^2$,0,-1,$\omega$), {\tt G}=(-1,-1,0,$\omega^2$), {\tt H}=(1,0,-1,$\omega^2$),\break {\tt I}=(-1,$\omega^2$,-1,0), {\tt J}=($\omega^2$,-1,1,0), {\tt K}=($\omega^2$,1,0,-1), {\tt L}=($\omega$,-1,1,0), {\tt M}=(1,0,0,0), {\tt N}=(1,1,0,-1), {\tt O}=(-1,-1,0,$\omega$), {\tt P}=($\omega$,-1,$\omega^2$,0), {\tt Q}=(1,-1,$\omega^2$,0), {\tt R}=($\omega^2$,0,-1,1), {\tt S}=(1,$\omega^2$,0,-1), {\tt T}=(0,1,1,$\omega^2$), {\tt U}=(0,1,$\omega$,1), {\tt V}=(1,-1,$\omega$,0), {\tt W}=(0,1,$\omega^2$,$\omega$), {\tt X}=($\omega^2$,$\omega$,0,-1), {\tt Y}=(0,1,1,$\omega$), {\tt Z}=(0,1,$\omega^2$,1), {\tt a}=(0,$\omega$,1,1), {\tt b}=(-1,0,$\omega^2$,-1), {\tt c}=(0,$\omega^2$,1,1), {\tt d}=(-1,0,$\omega$,-1), {\tt e}=($\omega$,0,-1,$\omega^2$)\}

Now, it was shown in \cite[Supplemental Material]{pwma-19} that
the only critical KS subgraphs of 40-40 are 40-25, 40-24, and
40-23. Hence, 28-14 (of \cite[TABLE V]{cabello-25}) cannot be a
critical KS set. Actually, it is not a contextual set at all.

{\bf 28-14} {\tt 1234,5678,639A,BCDE,2FGH,IGJK,L7JM,\break NOH8,1I5D,DHMA,PQ4M,RSKA,L9FE,4KE8.} \{{\tt 1}=(1,$\omega$,-1,0), {\tt 2}=(-1,-1,$\omega$,0), {\tt 3}=($\omega$,1,-1,0), {\tt 4}=(0,0,0,1), {\tt 5}=(-1,$\omega$,0,-1), {\tt 6}=($\omega$,-1,0,1), {\tt 7}=(1,-1,0,$\omega$), {\tt 8}=(0,0,1,0), {\tt 9}=($\omega$,0,1,-1), {\tt A}=(0,1,1,1), {\tt B}=($\omega^2$,0,$\omega$,-1), {\tt C}=($\omega$,0,$\omega^2$,-1), {\tt D}=(1,0,1,-1),\break {\tt E}=(0,1,0,0), {\tt F}=(1,0,$\omega$,-1), {\tt G}=(0,1,$\omega$,1), {\tt H}=(1,-1,0,1), {\tt I}=(0,$\omega$,1,1), {\tt J}=(0,1,1,$\omega$), {\tt K}=(1,0,0,0), {\tt L}=(-1,0,-1,$\omega$), {\tt M}=(1,1,-1,0), {\tt N}=($\omega$,-1,0,$\omega^2$), {\tt O}=($\omega^2$,-1,0,$\omega$), {\tt P}=($\omega$,$\omega^2$,-1,0), {\tt Q}=($\omega^2$,$\omega$,-1,0), {\tt R}=(0,1,$\omega$,$\omega^2$), {\tt S}=(0,1,$\omega^2$,$\omega$)\}

We can easily verify it on any PC as follows.

COMMAND-S: states01 -1 -c < {\bf 28-14}

OUTPUT: {\bf 28-14} fails (admits \{0,1\} state)

\subsection{\label{subsec:3dim01} IV F,G,H,I,J and
  TABLES VI, VII of \cite{cabello-25} ($\boxed{\ne}$,
  \faUnlink\ , \faHourglassHalf\ )}

$\boxed{\ne}$, \faUnlink\  ``Conway and Kochen’s
37-vector set in [67] has 22 orthogonal bases. It is an
extension of CK-31. The KS set \dots\ [is] shown in
Table VI.'' \cite[Sec.~IV G, p.~9]{cabello-25}
TABLE VI. `` The upper part of the table displays the 37
vectors of CK-37\dots CK-37 has 22 orthogonal bases'' 
\cite[TABLE VI, p.~9]{cabello-25}

\begin{figure}[ht]
\begin{center}
  \includegraphics[width=0.49\textwidth]{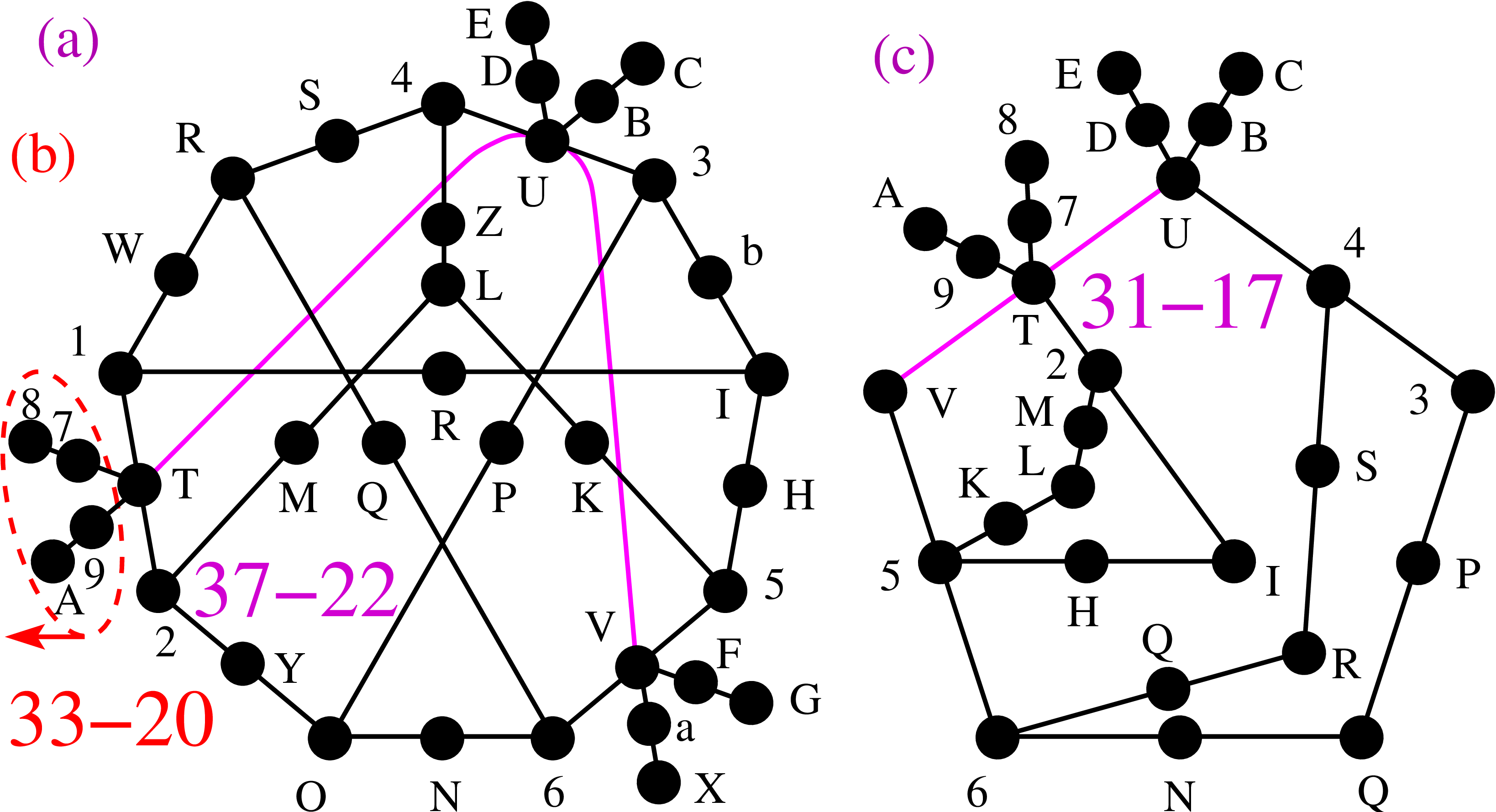}
\end{center}
\caption{(a) 37-22 set (CK-37); (b) 33-20 (CK-33) which
  is obtained from 37-22 by removing $m$-tuples {\tt 87T}
  and {\tt A9T}; (c) 31-17 set (CK-31); none of these sets
  is contextual since they all satisfy conditions (i) and
  (ii) of the KS Theorem \ref{theorem:KS}, i.e., admit \{0,1\}
  states; note that $m$-tuples attached to the main body of
  the sets by just one vector (e.g., {\tt UBC}) are
  structurally irrelevant and can be safely removed.}
\label{fig:CK}
\end{figure}

This is wrong:

{\parindent=0pt CK-37, i.e., {\bf 37-22} {\tt 34U,56V,FGV,3Ib,5HI,1IJ,4LZ,5KL,\break 2LM,3OP,6NO,2OY,4RS,6QR,1RW,12T,78T,9AT,TUV,\break BCU,DEU,VXa.} \{{\tt 1}=(1,1,0), {\tt 2}=(1,-1,0), {\tt 3}=(1,0,-1), {\tt 4}=(1,0,1), {\tt 5}=(0,1,1), {\tt 6}=(0,1,-1), {\tt 7}=(1,-2,0), {\tt 8}=(2,1,0), {\tt 9}=(1,2,0), {\tt A}=(2,-1,0), {\tt B}=(2,0,-1), {\tt C}=(1,0,2), {\tt D}=(1,0,-2), {\tt E}=(2,0,1), {\tt F}=(0,2,-1), {\tt G}=(0,1,2), {\tt H}=(2,1,-1), {\tt I}=(1,-1,1), {\tt J}=(-1,1,2), {\tt K}=(2,-1,1), {\tt L}=(1,1,-1), {\tt M}=(1,1,2), {\tt N}=(-2,1,1), {\tt O}=(1,1,1), {\tt P}=(1,-2,1), {\tt Q}=(2,1,1), {\tt R}=(-1,1,1), {\tt S}=(1,2,-1), {\tt T}=(0,0,1), {\tt U}=(0,1,0), {\tt V}=(1,0,0), {\tt W}=(1,-1,2), {\tt X}=(0,1,-2), {\tt Y}=(1,1,-2), {\tt Z}=(-1,2,1), {\tt a}=(0,2,1), {\tt b}=(1,2,1)\},

shown in Fig.~\ref{fig:CK}(a), is a noncontextual set:

COMMAND-S: states01 -1 < {\bf 37-22}

OUTPUT: {\bf 37-22} admits \{0,1\} state}

Therefore, CK-37 is neither a KS nor a non-KS state; it is
not a contextual set. There are no known 3-dim KS sets with
fewer than 36 bases. The 2nd part of TABLE VI is wrong.

\smallskip
$\boxed{\ne}$, \faUnlink\  ``Conway and Kochen’s
31-vector KS set [CK-31] in [67] is a critical KS set and
is currently the smallest \dots known KS set in dimension 3.
It has 17 orthogonal bases.''
\cite[Sec.~IV F, p.~8]{cabello-25} ``the KS set CK-31 is
obtained from CK-37 by removing vectors vectors
$v_{32},\dots,v_{37}$'' \cite[TABLE VI, p.~9]{cabello-25}

This is wrong.

{\parindent=0pt CK-31, i.e., {\bf 31-17} {\tt 34U,56V,FGV,5HI,1IJ,5KL,2LM,3OP,\break 6NO,4RS,6QR,12T,78T,9AT,TUV,BCU,DEU.}

shown in Fig.~\ref{fig:CK}(c), is a noncontextual set:
  
COMMAND-S: states01 -1 < {\bf 31-17}

OUTPUT: {\bf 31-17} admits \{0,1\} state}

Therefore, CK-31 is neither a KS nor a non-KS state; it is
not a contextual set. There are no known 3-dim KS sets with
fewer than 36 bases. The 4th part of TABLE VI is wrong. 

\smallskip
$\boxed{\ne}$, \faUnlink\  ``Conway and Kochen’s
33-vector KS set in [67], hereafter called CK-33
(or Sch\"utte-33 [68]), is also a critical KS set\dots
It has 20 orthogonal bases. Table VI illustrates CK-33.''

This is wrong.

{\parindent=0pt CK-33, i.e., {\bf 33-20} {\tt 34U,56V,FGV,3Ib,5HI,1IJ,4LZ,5KL,\break 2LM,3OP$\!$,6NO,2OY$\!$,4RS,6QR,1RW$\!$,12T$\!$,TUV,BCU,DEU,VXa.}

shown in Fig.~\ref{fig:CK}(b), is a noncontextual set:
  
COMMAND-S: states01 -1 < {\bf 33-20}

OUTPUT: {\bf 33-20} admits \{0,1\} state}

Therefore, CK-33 is neither a KS nor a non-KS state; it is
not a contextual set. There are no known 3-dim KS sets with
fewer than 36 bases. The 3rd part of TABLE VI is wrong. 

It is confusing that the authors did not notice that all
three of their sets (CK-31,33,37) contain $m$-tuples
attached to the main body of the sets by just one vector;
such $m$-tuples cannot contribute to contextually of the
sets and can therefore be safely removed
(see Fig.~\ref{fig:CK}). This is inherently included in
the very definition of MMP hypergraphs as ``no hyperedge
shares only one vertex with another hyperedge''
\cite{pavicic-entropy-25}.

\faUnlink\ , \faHourglassHalf\ ``In Ref. [52],
Man\v cinska constructed \dots 57-vector KS set (henceforth
MP-57) [with] 40 \dots bases \cite[Sec.~J, P10]{cabello-25}.
TABLE VII. MP-57.
\cite[Sec.~IV I, p.~10, TABLE VII, p.~11]{cabello-25}

The 57-40 Peres' 3-dim KS set (isomorphic to MP-57) was
explicitly given 20 years ago in
\cite[Footnote 9, p.~1590]{pmmm05a}.
As explained in
\cite[4th par., left column, p.~3]{pavicic-pra-22}
$\{0,\pm1,\pm\sqrt{2},3\}$ generates the master set 81-52 
via COMMAND-M. This master set contains only one critical KS
set and that is Peres' 57-40 obtained via COMMAND-S, all within
seconds on a PC: 

{\bf 57-40} {\tt 123,147,456,5QR,6ST,789,8MN,9OP,ATZ,1AB,\break BNa,BOb,1CD,CRc,CSd,DMe,DPf,4EF,FMo,FOp,4GH,\break HNq,HPr,7IJ,JSu,7KL,LRt,LTv,AQY,JQs,EUg,KUh,\break 2UV,GVi,IVj,GWk,KWl,3WX,EXm,IXn.} \{{\tt 1}=(0,0,1), {\tt 2}=(1,-1,0), {\tt 3}=(1,1,0), {\tt 4}=(0,1,0), {\tt 5}=(1,0,-1), {\tt 6}=(1,0,1),\break {\tt 7}=(1,0,0), {\tt 8}=(0,1,1), {\tt 9}=(0,1,-1), {\tt A}=($\sqrt{2}$,-1,0), {\tt B}=(1,$\sqrt{2}$,0),\break {\tt C}=($\sqrt{2}$,1,0), {\tt D}=(1,-$\sqrt{2}$,0), {\tt E}=($\sqrt{2}$,0,-1), {\tt F}=(1,0,$\sqrt{2}$),\break {\tt G}=($\sqrt{2}$,0,1), {\tt H}=(1,0,-$\sqrt{2}$), {\tt I}=(0,$\sqrt{2}$,1), {\tt J}=(0,1,-$\sqrt{2}$), {\tt K}=(0,$\sqrt{2}$,-1), {\tt L}=(0,1,$\sqrt{2}$), {\tt M}=($\sqrt{2}$,1,-1), {\tt N}=($\sqrt{2}$,-1,1), {\tt O}=(-$\sqrt{2}$,1,1), {\tt P}=($\sqrt{2}$,1,1), {\tt Q}=(1,$\sqrt{2}$,1), {\tt R}=(1,-$\sqrt{2}$,1), {\tt S}=(-1,$\sqrt{2}$,1), {\tt T}=(1,$\sqrt{2}$,-1), {\tt U}=(1,1,$\sqrt{2}$), {\tt V}=(1,1,-$\sqrt{2}$), {\tt W}=(-1,1,$\sqrt{2}$), {\tt X}=(1,-1,$\sqrt{2}$), {\tt Y}=(-1,-$\sqrt{2}$,3), {\tt Z}=(1,$\sqrt{2}$,3), {\tt a}=(-$\sqrt{2}$,1,3), {\tt b}=($\sqrt{2}$,-1,3), {\tt c}=(-1,$\sqrt{2}$,3), {\tt d}=(1,-$\sqrt{2}$,3),\break {\tt e}=($\sqrt{2}$,1,3), {\tt f}=(-$\sqrt{2}$,-1,3), {\tt g}=(-1,3,-$\sqrt{2}$), {\tt h}=(3,-1,-$\sqrt{2}$),\break {\tt i}=(-1,3,$\sqrt{2}$), {\tt j}=(3,-1,$\sqrt{2}$), {\tt k}=(1,3,-$\sqrt{2}$), {\tt l}=(3,1,$\sqrt{2}$), {\tt m}=(1,3,$\sqrt{2}$), {\tt n}=(3,1,-$\sqrt{2}$), {\tt o}=(-$\sqrt{2}$,3,1), {\tt p}=($\sqrt{2}$,3,-1), {\tt q}=($\sqrt{2}$,3,1), {\tt r}=(-$\sqrt{2}$,3,-1), {\tt s}=(3,-$\sqrt{2}$,-1), {\tt t}=(3,$\sqrt{2}$,-1), {\tt u}=(3,$\sqrt{2}$,1),\quad{\tt v}=(3,-$\sqrt{2}$,1)\}

That means that the vectors in TABLE VII obtained from vector
components
$\{0,\pm1,\pm\sqrt{2},\pm1/3,\pm3/\sqrt{2},\pm/\sqrt{2},\pm3\}$
are overcomplicated. 

$\boxed{\ne}$, \faUnlink\ , \faHourglassHalf\
``Sec. IV. I. P-33. Peres’ 33-vector set in d = 3 [62],
hereafter called P-33 is a critical KS set. It has 16
orthogonal bases. The KS set [is] illustrated in TABLE VII''
\cite[pp.~10,11]{cabello-25}.

This is doubly wrong. First, it is not critical. See also
Ref.~\cite{pavicic-entropy-23}. COMMAND-S: 

{\parindent=0pt {\bf 33-40} fails (not critical):
{\tt 123,147,456,5QR,6ST,789,8MN,\break 9OP,AT,1AB,BN,BO,1CD,CR,CS,DM,DP,4EF,FM,FO,4GH,\break HN,HP,7IJ,JS,7KL,LR,LT,AQ,JQ,EU,KU,2UV,GV,IV,\break GW,KW,3WX,EX,IX.}}

Second, as it is obvious from this string and the 3rd part of
TABLE VII, it does not have 16, but 40 bases. 

The second erroneous statement might be a misprint, but the
1st one is surprising since the authors should be aware that
33-40 cannot be a critical KS set as shown in
\cite{pavicic-entropy-23}.  

Moreover, none of the smallest 3-dim non-KS sets,
Bub's (Sch\"utte's) 33-36, Conway-Kochen's 31-37,
Peres' 33-47, Kochen-Specker's 117-118, is critical
and all of them contain abundant distribution of
smaller non-KS sets, the smallest of which are
5-5 and 8-7 \cite[Fig.~4,p.~4]{pavicic-entropy-19}

TABLE VI gives a coordinatization of CK-31, CK-37, CK-33
which is generated from vector components $\{0,\pm1,\pm2\}$.
However, in order to generate 57-37 or 49-36 so as to be
implementable in a lab one needs $\{0,\pm1,\pm2,5\}$.

\subsection{\label{subsec:8dim01} IV K,L and
  TABLE VIII of \cite{cabello-25}
  (\faHourglassHalf\ )}

\faHourglassHalf\ ``Kernaghan and Peres’ 36-vector KS set
in d=8 [71] hereafter called KP-36, is a critical KS set.''
\cite[Sec.~K, p.~10, TABLE VIII, p.~12]{cabello-25}
``Kernaghan and Peres’ 40-vector KS set in d=8 [71],
hereafter called KP-40, has 25 orthogonal bases'' 
\cite[Sec.~L, p.~10, TABLE VIII, p.~12]{cabello-25}

It is rather seclusive to involve these two KS sets from
a 30 years old paper when 8 years ago 6,925,540 critical
8-dim KS sets were generated, the smallest of which are
two nonisomorphic 34-9 criticals:

{\parindent=0pt {\bf 34-9(a)} {\tt 12345678,9ABCDEFG,HIJKLMFG,NOPQME78,\break RSTULD56,VWXUKC46,XTPQJB36,YVWOIA28,YRSNH918.}

  \smallskip
  {\bf 34-9(b)} {\tt 12345678,9ABC5678,DEFG3478,HIJKLMFG,\break NOPQRMEC,STUQRLDB,VWUPJK9A,XYWTOI28,XYVSNH17.} 

\smallskip
and eight nonisomorphic 36-9 criticals
\cite[Fig.~13, p.~15; Fig.~14, p.~16]{pavicic-pra-17}}

They all belong to the 3280-1361376 master generated from
vector components $\{0,\pm1\}$
\cite[Table 2, p.~6]{pm-entropy18}. 

There is a misprint in the caption of
\cite[TABLE VIII, p.~12]{cabello-25}:
``KP-40 has \dots 24 \dots bases, numbered from 1 to 24.''
It is not 24, but 25.

\subsection{\label{subsec:5dim01} IV M,N and
  TABLE IX of \cite{cabello-25}
  (\faHourglassHalf\ )}

\vspace{-10pt}

\faHourglassHalf\ ``The KS set S-29 is built from CEG-18
via a recursive construction introduced in Ref. [72].
'' \cite[Sec.~M, p.~10, TABLE IX, p.~12]{cabello-25}

It is again seclusive to refer to a single KS set obtained
20 years ago, when 14 months ago 28 million critical 5-dim
KS sets were generated which include S-29 (29-16).
\cite[Fig.~2, p.~4]{pavicic-pra-22}

\vspace{-10pt}

\subsection{\label{subsec:7dim01} IV O,P and
  TABLE X of \cite{cabello-25}
  (\faHourglassHalf\ )}

\vspace{-10pt}

\faHourglassHalf\ ``The KS set in dimension 7 with fewest
number of vectors known is S-34, which was introduced in
Ref. [60]. It has 28 bases and is critical.''
\cite[Sec.~O), p.~10, TABLE X, p.~13]{cabello-25}

Yet another ancient reference to a single KS set obtained 30
years ago, when a year ago 42,816 critical 7-dim KS 
sets were generated by one method
\cite[Fig.~3, p.~4]{pavicic-pra-22} and a million by another
\cite[Fig.~4, p.~5; Appendix 4, p.~10]{pw-23a}. They both
include 34-14 (half of the bases S-34 has).  

\vspace{-10pt}

\section{\label{subsec:concl} Conclusion}

Quidquid agis prudenter agas et respice finem. Festina lente. 

\section{\label{subsec:cnote} Note}

Reports on the previous version of the paper as well as my
responses to them and a list of changes I made in the present
version of the paper are included in the ancillary file. 

\vbox to 10pt {}

{\em Acknowledgements}---Supported by the Ministry of Science and
Education of Croatia through Center of Excellence CEMS funding,
and by the MSE grant No.~KK.01.1.1.01.0001. Computational support
was provided by the Zagreb University Computing Centre.
Programs are freely available from our repository
\cite{puh-repository}.

\begin{widetext}

\section{Appendix: Response to Reply to Pavi\v ci\'c's `Comment on ``Optimal conversion of {K}ochen-{S}pecker sets into bipartite
  perfect quantum strategies'' (arXiv:2502.13787)'}

  The main point of my Comment is that a number of statements
  Trandafir and Cabello made in their paper \cite{cabello-25}
  are ``wrong.'' In their Reply \cite{cabello-25b} the authors
  claim that the statements are wrong in ``my'' definition of a
  Kochen-Specker (KS) set, but not in ``their ``original''''
  definition of a KS set. However, in my Comment I show (i) that
  the statements are wrong because they are noncontextual and
  therefore are not KS sets by any definition, while Trandafir
  and Cabello nevertheless call them ``KS sets,'' i.e., label
  them contextual; (ii) that ``my'' and ``their'' definitions
  of KS-like sets coincide---they are just ``cross-named.'' 
  Here, I provide some more details about all that since the
  authors do not discuss their errors any further in their
  Ref.~\cite{cabello-25b}  and since they misunderstood several
  further points I put forward in the Comment. The bulk of their
  \cite{cabello-25b} (their Theorems 1, 2, 2b, and 2c) represents
  another misunderstanding which I nevertheless discuss below. 
\end{widetext}

   \setcounter{section}{0}

 \section{\label{sec:intro}Kochen-Specker set
            definitions}

In my Comment I introduce definitions of a KS and non-KS set.

\begin{theorem}\label{theorem:KS}
  {\bf KS (non-KS)}
  {\rm \cite{zimba-penrose,renner-wolf-04,pmmm05a,pavicic-quantum-23,pavicic-entropy-25}.}
  In ${\cal H}^n$, $n\ge 3$, there are sets, called KS
  (non-KS) sets, of $n$-tuples ($m$-tuples, $2\le m\le n$)
  of mutually orthogonal vectors to which it is impossible
  to assign $1$s and $0$s so that

{\parindent=0pt
(i)  No two orthogonal vectors are both assigned the value $1$;

(ii) In no $n$-tuple ($m$-tuple) of mutually orthogonal
  vectors, all of the vectors are assigned the value $0$.}
\end{theorem}

The authors claim that ``In all cases, Pavi\v ci\'c
\cite{pavicic-arxiv-25-v1} assumes a different definition of a
KS set'' and that their definition of a KS set runs as
follows---it is actually our ``non-KS'' set:

\setlength{\leftskip}{20pt}

{\parindent=0pt
  ``{\em Definition.\/}---A KS set \dots is a finite
  set \dots which does not admit an assignment $f:V\to\{0,1\}$
  satisfying: (I) $f(u)+f(v)\le 1$ for each pair of orthogonal
  [vectors] $u,v\in V$; (II) $\sum_{u\in b}=1$ for every set
  $b\subset V$ of mutually orthogonal [vectors] whose sum is the
  identity.''}

\setlength{\leftskip}{0pt}

For what we call a KS set they use the term ``extended'' KS set:

\setlength{\leftskip}{20pt}

{\parindent=0pt
``an alternative definition in which condition (I) is removed,
Larsson refers to as ``extended'' KS sets.''}

\setlength{\leftskip}{0pt}

\smallskip 
Hence, the definitions are the same, just ``cross-named.''

{\parindent=0pt
\framebox[1.04\width]{Our KS sets are their ``extended'' KS sets
   and vice versa.}

and

\framebox[1.04\width]{Our non-KS sets are their KS sets and
   vice versa.}}

\medskip 
The authors write that ``[my claims that their statements about
their sets are ``wrong''] are based on the fact that [I am]
assuming a different definition of KS set.'' 

\medskip

{\parindent=0pt
\framebox[1.02\width]{
  \begin{minipage}{238pt}
  However, what is ``wrong'' with their sets is that they
  satisfy conditions (i) and (ii) of my definitions and (I) and
  (II) of theirs, i.e., that they do admit assignment
  $f:V\to\{0,1\}$, i.e., that they are noncontextual, i.e.,
  that they are \_not\_ contextual, while the authors
  call them ``KS sets'' and, in effect, falsely label them
  contextual.
  \end{minipage}}

\medskip
\framebox[1.02\width]{
  \begin{minipage}{238pt}
    It is ``wrong'' to call a noncontextual
    set either a ``KS set,'' or a ``non-KS set,'' or an
    ``extended KS set;'' it is ``wrong'' to label it
    ``contextual.''
  \end{minipage}}}

\bigskip
I elaborate on all ``wrong'' statements in the Comment but
let me sum up the elaboration in Secs.~\ref{sec:err1} and
\ref{sec:err3}.

Thereafter, in Sec.~\ref{sec:err2}, I discuss differences
between two types of KS sets the authors elaborated on in
\cite{cabello-25b} while commenting on my characterization
of their usage.

\vspace{-10pt}
\section{\label{sec:err1} ``Wrong'' 
  statements $\boxed{\ne}$}. 

\vspace{-20pt}

Notice that all sets in this section have only full triples
(each triple has three vectors) and that none of the conditions of
the KS theorem \ref{theorem:KS} are violated. Hence, the considered
sets are neither KS nor non-KS nor ``extended'' KS sets. They all
allow 0-1 assignments and are therefore noncontextual.

\vspace{-10pt}

\subsection{\label{subsec:err11}
  ZP-28 is not a contextual set $\boxed{\ne}$} 

\vspace{-10pt}
Trandafir and Cabello \cite[p.~8, left column---bottom]{cabello-25}
and \cite[TABLE V, p.~9]{cabello-25}:

\setlength{\leftskip}{20pt}

{\parindent=0pt

$\boxed{\ne}$ ``Zimba-Penrose’s 28-vector KS set [ZP-28] \dots is a
critical KS set formed by removing 12 implicit vectors of Pen-40.''}

\setlength{\leftskip}{0pt}
In my Comment, Sec.~B, ZP-28 is extracted from the
TABLE IV of the authors and presented explicitly together with its
coordinatization (not provided by the authors). It is a 28-14 set
which reads: {\bf 28-14} {\tt 1234,5678,\dots}
{\tt 1}=(1,$\omega$,-1,0),\break 
{\tt 2}=(-1,-1,$\omega$,0), {\tt 3}=($\omega$,1,-1,0),\dots

It is provably noncontextual:

COMMAND-S: states01 -1 -c < {\bf 28-14}

OUTPUT: {\bf 28-14} fails (admits \{0,1\} assignment)

This can easily be verified on any PC.
  
{\parindent=0pt
\framebox[1.02\width]{
  \begin{minipage}{238pt}
    $\boxed{\ne}$ Hence, their statement ``ZP-28 is a critical KS
    set'' is wrong. ZP-28 is not a contextual set. ZP-28 is not a
    KS set of any kind.
\end{minipage}}}

\vspace{-10pt}

\subsection{\label{subsec:err12} CK-37 is not a contextual  
  set $\boxed{\ne}$}

\vspace{-10pt}
Trandafir and Cabello \cite[Sec.~IV G, p.~9]{cabello-25}
and \cite[TABLE VI, p.~9]{cabello-25}:

\setlength{\leftskip}{20pt}

{\parindent=0pt
  $\boxed{\ne}$, ``Conway and Kochen’s 37-vector set in
  \cite{peres-book} has 22 orthogonal bases. It is an extension
  of CK-31. KS [CK-37] set \dots\ shown in Table VI. \dots has
  22 orthogonal bases''}

\setlength{\leftskip}{0pt}

In my Comment, Sec.~C, CK-37 is extracted from TABLE VI of the
authors and presented explicitly together with its coordinatization:
{\bf 37-22} (CK-37) {\tt 34U,56V,FGV,3Ib,\dots} {\tt 1}=(1,1,0),
{\tt 2}=(1,-1,0), \dots I even show its hypergraph in Fig.~1(a).

It is provably a noncontextual set:

COMMAND-S: states01 -1 < {\bf 37-22}

OUTPUT: {\bf 37-22} admits \{0,1\} assignment

Therefore, CK-37 is neither a KS nor a non-KS set; it is
not a contextual set. There are no known 3-dim KS sets with
fewer than 36 bases. The 2nd part of TABLE VI is wrong. 
\framebox[1.05\width]{
  \begin{minipage}{230pt}
 $\boxed{\ne}$ Hence, their statement ``CK-37 is a KS set'' is wrong.
 CK-37 is not a KS set of any kind; it is noncontextual.
\end{minipage}}

\vspace{-20pt}

\subsection{\label{subsec:err13} CK-31 is not a contextual  
  set $\boxed{\ne}$}

\vspace{-10pt}

Trandafir and Cabello \cite[Sec.~IV F, pp.~8,9]{cabello-25}
and \cite[TABLE VI, p.~9]{cabello-25} write: 

\setlength{\leftskip}{20pt}

{\parindent=0pt
  $\boxed{\ne}$ ``Conway and Kochen’s 31-vector KS set \dots
  is a critical KS set.\dots It has 17 orthogonal bases.''}

\setlength{\leftskip}{0pt}

In my Comment, Sec.~C, CK-31 is extracted from TABLE VI of the
authors:: {\bf 31-17} (CK-31) {\tt 34U,56V,FGV,\dots,DEU.} and
its hypergraph is shown in Fig.~1.(c)

It is provably a noncontextual set:

COMMAND-S: states01 -1 < {\bf 31-17}

OUTPUT: {\bf 31-17} admits \{0,1\} assignment

Therefore, CK-31 is neither a KS nor a non-KS aet; it is
not a contextual set. There are no known 3-dim KS sets with
fewer than 36 bases. The 2nd part of TABLE VI is wrong. 
\framebox[1.05\width]{
  \begin{minipage}{230pt}
 $\boxed{\ne}$ Hence, their statement ``CK-31 is a critical KS
 set'' is wrong. CK-31 is not a KS set of any kind; it is
 noncontextual. And it is not Conway-Kochen's 31-vector set. 
\end{minipage}}

\vspace{-20pt}

\subsection{\label{subsec:err14} CK-33 is not a contextual  
  set $\boxed{\ne}$}

\vspace{-10pt}

Trandafir and Cabello \cite[Sec.~IV H, p.~9]{cabello-25}
and \cite[TABLE VI, p.~9]{cabello-25} write: 

\setlength{\leftskip}{20pt}

{\parindent=0pt
  $\boxed{\ne}$ ````Conway and Kochen’s
  33-vector KS set \cite{peres-book},\dots CK-33 \dots is a critical KS
  set\dots It has 20 orthogonal bases.''}

  \setlength{\leftskip}{0pt}

In my Comment, Sec.~C, CK-33 is extracted from the TABLE VI of the
authors:: {\bf 33-20} (CK-33) {\tt 34U,56V,FGV,\dots,VXa.} and its
hypergraph is shown in Fig.~1.(b)

It is provably a noncontextual set:

COMMAND-S: states01 -1 < {\bf 33-20}

OUTPUT: {\bf 33-20} admits \{0,1\} assignment

Therefore, CK-33 is neither a KS nor a non-KS set, let alone a
critical KS set; it is not a contextual set. There are no known
3-dim (extended) KS sets with fewer than 36 bases. The 2nd part
of TABLE VI is wrong. 
\framebox[1.05\width]{
  \begin{minipage}{230pt}
 $\boxed{\ne}$ Hence, their statement ``CK-33 is a critical KS
 set'' is wrong. CK-33 is not a KS set of any kind; it is
 noncontextual. 
\end{minipage}}

\vspace{-15pt}

\subsection{\label{subsec:err15} Ergo, claim on
  ``wrong'' statements in \cite{cabello-25b} is wrong. }

\vspace{-15pt}

Trandafir and Cabello in their \cite{cabello-25b} write: 

\setlength{\leftskip}{20pt}

{\parindent=0pt
  $\boxed{\ne}$ ``There is nothing wrong [with our statements]\dots
  Pavi\v ci\'c is assuming a different definition of [a KS set].''}

  \setlength{\leftskip}{0pt}

{\parindent=0pt
\framebox[1.02\width]{
  \begin{minipage}{238pt}
  As it is clear from the previous
  Subsecs.~\ref{subsec:err11}--\ref{subsec:err14}, their ``wrong''
  statements on their sets have nothing to do with the definition
  of a KS set since their sets are noncontextual, i.e., are not
  KS sets by any definition, i.e., all is ``wrong.''.
\end{minipage}}}

\vspace{-5pt}

\section{\label{sec:err3}Another ``wrong'' statement}

\vspace{-5pt}

Trandafir and Cabello \cite[Sec.~IV I, p.~10]{cabello-25}
and \cite[TABLE VII, p.~11]{cabello-25} write:

\setlength{\leftskip}{20pt}

{\parindent=0pt
  \faUnlink\  ``Peres’ 33-vector set in d=3 \cite{peres}, hereafter
  called P-33 is a critical KS set. It has 16 orthogonal bases.''}

  \setlength{\leftskip}{0pt}

{\parindent=0pt
\framebox[1.02\width]{
  \begin{minipage}{238pt}
This is doubly wrong. The set P-33 is 33-40. First, it is not
  critical. Second, it does not have 16, but 40 bases.
\end{minipage}}}

\smallskip  

  COMMAND-S: states01 -1 < 33-40

  OUTPUT: fails (not critical).

  See also Ref.~\cite{pavicic-entropy-23}.

\vspace{-5pt}

\section{\label{sec:err2}Consistency --- Our
  $\rm vs.$~their $\rm vs.$~non- $\rm
  vs.$~extended KS sets}

\vspace{-5pt}

Trandafir and Cabello in their \cite{cabello-25b} write: 

\setlength{\leftskip}{20pt}

{\parindent=0pt
  \faUnlink\  ``Larsson \cite{larsson}, introduced an alternative
  definition in which condition (I) is removed. Larsson refers
  to the resulting sets as “extended” KS sets \cite{larsson},
  since they are constructed by adding observables to the original
  KS sets. Every extended KS set is an original KS set, but not
  the other way around.''}

  \setlength{\leftskip}{0pt}

This is inconsistent (\faUnlink\ ) for two reasons:

First, Larsson does not ``add'' vectors, he ``recovers'' them:  
``in some [Peres' 33] cases {\em a rotation goes from one of the
triads to a pair of vectors where the third vector needed to form a
triad is not in the set}\dots [So,] extra vectors are needed [and]
the total number of added vectors is 24, yielding [the set 57-40].''
\cite[p.~804]{larsson}. 

Second, both Cabello and Larsson are coauthors of a paper
in {\em Rev.~Mod.~Phys.} where they write:
``For any two vectors that are orthogonal but do not [form] 
a basis (hyperedge), one can readily add a vector to complete the
pair to a basis. This enlarges the [original Kochen-Specker's] 117
vectors to 192 vectors and those 192 vectors
form then the KS set in the KS theorem''
\cite[p.~6]{budroni-cabello-rmp-22}.

{\parindent=0pt
\framebox[1.02\width]{
  \begin{minipage}{238pt}
    \faUnlink\  So, Cabello apparently seems to contradict
  himself when claiming that ``every extended KS set is an original
  KS set.''
\end{minipage}}}

\bigskip
Yet, Trandafir and Cabello in their \cite{cabello-25b}, 
write:  

\setlength{\leftskip}{20pt}

{\parindent=0pt
  ``Beyond terminology, there are fundamental physical reasons
for focusing on the original KS sets. One reason is that,
for experimentally observing quantum state-independent
contextuality, there is no need to measure the observables added
in the extended KS sets; the observables in the original KS
sets suffice. ''}

  \setlength{\leftskip}{0pt}

\smallskip

{\parindent=0pt
\framebox[1.01\width]{
    \begin{minipage}{240pt}
Whether we measure an outcome or not is not a
``fundamental physical reason''; the outcome exists no matter
whether we measure it or not. Instead, fundamental physical
reasons for considering ``extended'' KS sets might be: 
\end{minipage}}}

\smallskip
``In any KS diagram and therefore in Kochen–Specker 192-118,
Peres 57-40, and Conway–Kochen’s 51-37, only all vectors together
make a complete description of their KS sets. Recall that one
arrives from an $n$-tuple to an adjoining one by rotation around
an $(n-2)$-dimensional subspace'' \cite{pmmm05a}.

The crucial point is that ``extended'' KS sets Kochen–Specker
192-118, Peres 57-40, and Conway–Kochen’s 51-37 are all critical
KS sets meaning that a removal of any of their hyperedges turn
them into noncontextual (0-1 colorable) sets. In contrast,
none of Kochen–Specker 117-118, Peres 33-40, and Conway–Kochen’s
31-37 is critical, meaning that by removing their hyperedges we
arrive at smaller contextual sets \cite{pavicic-entropy-23}.

\vspace{-10pt}

\section{\label{sec:eff}Efficiency}

\vspace{-7pt}

Trandafir and Cabello in their \cite{cabello-25b}, p.~3, write: 

\smallskip
\setlength{\leftskip}{20pt}

{\parindent=0pt
  ``Pavi\v ci\'c also states that our representation of the KS
  sets in \cite{cabello-25} is ``inefficient'' [\faHourglassHalf].
  Our aim in \cite{cabello-25} was to provide information in a
  clear readable manner.''}

  \setlength{\leftskip}{0pt}

\medskip  
{\parindent=0pt
\framebox[1.02\width]{
  \begin{minipage}{238pt}
 \faHourglassHalf\ That is a problem. The authors do not
 grasp that their presentation is far from a ``clear
 readable manner''---it is simply outdated. A direct piping
 into a computer program is today's ``readable manner.'' And
 there is no indication that their information satisfy this
 requirement. 
\end{minipage}}}

\medskip  
Their outputs do not even enable a direct copy-paste into a
program. Is the reader expected to do that by hand as the
authors of their sources did 30 years ago?

Trandafir and Cabello in their \cite{cabello-25b}, p.~3, then
write: 

\smallskip
\setlength{\leftskip}{20pt}

{\parindent=0pt
  ``In most cases, we explicitly gave the vectors and orthogonal
  bases (as it is a standard practice
  \cite{bub,peres,peres-book}).''}

      \setlength{\leftskip}{0pt}

\medskip  
{\parindent=0pt
\framebox[1.02\width]{
  \begin{minipage}{238pt}
It {\em was} ``a standard practice'' 40 to 30 years ago when the
sources \cite{bub,peres,peres-book} were published. Today we are
able to provide readers with an automated procedure and language
of handling, generating, and processing of all possible KS sets
in any dimension by means of available computer programs. 
\end{minipage}}}

\medskip  

Trandafir and Cabello mislead the readers that they should go
through outdated papers, decades old, to find their KS inputs
instead of offering them up-to-date computer commands which
generate all KS sets they used in their paper on any PC in
no time. 

Trandafir and Cabello in their \cite{cabello-25b}, p.~3, in the
end, write: 

\smallskip
\setlength{\leftskip}{20pt}

{\parindent=0pt
  ``In addition, in Secs. E and F of \cite{pavicic-arxiv-25-v1},
  Pavi\v ci\'c provides references for KS sets\dots However,
  so far, all the records in dimensions from 3 to 8 are in
  \cite{cabello-25}.''}
 
      \setlength{\leftskip}{0pt}

\medskip  
{\parindent=0pt
\framebox[1.02\width]{
  \begin{minipage}{238pt}
 However, as we have seen above,
 in Subsecs.~\ref{subsec:err11}--\ref{subsec:err14},
 half of them are wrong. 
\end{minipage}}}

%\phantom{X}

Besides, referring to ``record small'' KS sets today is of no
relevance since computer programs can generate any KS of any
size in any dimension in no time.

\phantom{X}
\vspace{-7pt}

\section{\label{subsec:concl} Conclusion}

\vspace{-10pt}
In summary, I did my best to explain to Trandafir and Cabello
what their errors, inconsistencies, and inefficiencies are.

So, my claims about their errors do not seem to be disputable
and those about inconsistencies and inefficiencies are
apparently well founded.  

I recommend that the authors use our programs to verify the
errors they made.

If they still do not trust our algorithms and programs, they
should verify their outcomes manually.

%\vbox to 5pt {}

Programs are freely available from our
repository \cite{puh-repository}.


%\bibliography{/3rd-disk/ql/m-p.bib}

\end{document}